\documentclass[12pt]{article}

\usepackage[a4paper,left=30mm,right=20mm,top=20mm,bottom=20mm]{geometry}
\usepackage{graphics}
\usepackage{amsmath}
\usepackage{epsfig}
\usepackage{cite}
\usepackage{xcolor}

\newcommand{\z}{&&\hspace*{-1cm}}

\newcommand{\bea}{\begin{eqnarray}}
\newcommand{\eea}{\end{eqnarray}}
\newcommand{\be}{\begin{equation}}
\newcommand{\ee}{\end{equation}}

\newcommand{\ar}{a_s}

\title{About Fractional Analytic QCD}

\author{A.V.~Kotikov$^{1}$, I.A.~Zemlyakov$^{1,2}$}

\begin{document}

\maketitle

\begin{center}
 
{\it $^{1}$Joint Institute for Nuclear Research, 141980, Dubna, Moscow region, Russia}\\
{\it $^{2}$Dubna State University,
Dubna, Moscow Region, Russia}

\end{center}

\vspace{0.5cm}

\begin{center}

{\bf Abstract }

\end{center}

We present a brief  overview of fractional analytic QCD, basically following the results recently obtained in
Refs. \cite{Kotikov:2022sos,KoZe23}.\\


\label{sec:intro}
\section{Introduction}
According to the general principles of (local) quantum field theory (QFT) \cite{Bogolyubov:1959bfo},
observables in a spacelike region (i.e. in Euclidean space) can have singularities only for negative values of their argument $Q^2$.
However, for large $Q^2$ values, these observables are usually represented as power expansions in the running coupling constant (couplant)
$\alpha_s(Q^2)$,
which has a ghostly singularity, the so-called Landau pole, at $Q^2 = \Lambda^2$. Therefore, to restore the analyticity of the considered expansions,
this pole in the strong couplant should be removed.

The strong couplant $\alpha_s(Q^2)$ obeys  the renormalization group equation
\be
L\equiv \ln\frac{Q^2}{\Lambda^2} = \int^{\overline{a}_s(Q^2)} \, \frac{da}{\beta(a)},~~ \overline{a}_s(Q^2)=\frac{\alpha_s(Q^2)}{4\pi}\,
\label{RenGro}
\ee
with some boundary condition and the QCD $\beta$-function:
\be
\beta(\ar) ~=~ -\sum_{i=0} \beta_i \overline{a}_s^{i+2} 
=-\beta_0  \overline{a}_s^{2} \, \Bigl(1+\sum_{i=1} b_i \ar^i \Bigr),~~ b_i=\frac{\beta_i}{\beta_0^{i+1}}\,, ~~
\ar(Q^2)=
\beta_0\,\overline{a}_s(Q^2)\,, 
\label{beta}
\ee
where
\be
\beta_0=11-\frac{2f}{3},~~\beta_1=102-\frac{38f}{3},~~\beta_2=\frac{2857}{2}-\frac{5033f}{18}+\frac{325f^2}{54},~~
\label{beta_i}
\ee
for $f$ active quark flavors. Really now the first fifth coefficients, i.e. $\beta_i$ with $i\leq 4$, are exactly known \cite{Baikov:2016tgj}.
In our present consideration we will need only $0 \leq i\leq 2$.

Note that in Eq. (\ref{beta})
we have added the first coefficient of the  QCD $\beta$-function to the $\ar$ definition, as is usually done in the case of  
analytic couplants (see, e.g., Refs. \cite{ShS}-\cite{Bakulev:2010gm}).

So, already at leading order (LO), where $\ar(Q^2)=\ar^{(1)}(Q^2)$, we have from Eq. (\ref{RenGro})
\be
\ar^{(1)}(Q^2) = \frac{1}{L}\, ,
\label{asLO}
\ee
i.e. $\ar^{(1)}(Q^2)$ does contain a pole at $Q^2=\Lambda^2$.

In a timelike region  ($q^2 >0$) (i.e., in Minkowski space), the definition of a running couplant turns out to be quite difficult.
The reason for the problem is that, strictly speaking, the expansion of perturbation theory (PT) in QCD cannot be defined directly in
this region.
Since the early days of QCD, much effort has been made to determine the appropriate  Minkowski coupling parameter needed to describe
important timelike processes such as, $e^+e^-$-annihilation into hadrons, quarkonia and $\tau$-lepton decays into hadrons. Most of the
attempts (see, for example, \cite{Pennington:1981cw}) have been based on the analytical continuation of strong couplant from the deep Euclidean region,
where perturbative QCD calculations can be performed, to the Minkowski space, where physical measurements are made.
In other developments, analytical expressions for a LO couplant  were obtained \cite{Krasnikov:1982fx} directly in Minkowski space,
using an integral transformation from the spacelike to the timelike mode from the Adler D-function.

In Refs. \cite{ShS,MSS} an efficient approach was developed to eliminate the Landau singularity without introducing extraneous infrared controllers,
such as the gluon effective mass
(see, e.g., \cite{GayDucati:1993fn}).
This method is based on a dispersion relation that relates the new analytic couplant $A_{\rm MA}(Q^2)$ to the spectral function $r_{\rm pt}(s)$
obtained in the PT framework.
In LO this gives
    \be
A^{(1)}_{\rm MA}(Q^2) 
= \frac{1}{\pi} \int_{0}^{+\infty} \, 
\frac{ d s }{(s + t)} \, r^{(1)}_{\rm pt}(s),~~ r^{(1)}_{\rm pt}(s)= {\rm Im} \; a_s^{(1)}(-s - i \epsilon) \,.
\label{disp_MA_LO}
\ee
The \cite{ShS,MSS} approach follows the corresponding results \cite{Bogolyubov:1959vck} obtained in the framework of Quantum Electrodynamics.
  Similarly, the analytical images of a running coupling in the Minkowski space are defined using another linear operation
    \be
U^{(1)}_{\rm MA}(s) 
= \frac{1}{\pi} \int_{s}^{+\infty} \, 
\frac{ d\sigma }{\sigma} \, r^{(1)}_{\rm pt}( \sigma) \, ,
\label{disp_MAt_LO}
\ee

So, we repeat once again: the spectral function in the dispersion relations (\ref{disp_MA_LO}) and (\ref{disp_MAt_LO}) is taken directly from PT,
and the analytical couplants $A_{\rm MA}(Q^2)$ and $U_{\rm MA}(Q^2)$ are restored using the corresponding dispersion relations. This approach is usually
called the {\it Minimal Approach} (MA) (see, e.g., \cite{Cvetic:2008bn})
or the {\it Analytical Perturbation Theory} (APT) \cite{ShS,MSS}.
\footnote{An overview of other similar approaches can be found in \cite{Bakulev:2008td}, including approaches \cite{Nesterenko:2003xb} that are close to APT.}

Thus, MA QCD is a very convenient approach that combines the analytical properties of QFT quantities and the results
obtained in the framework of perturbative QCD, leading to the appearance of the MA couplants $A_{\rm MA}(Q^2)$ and $U_{\rm MA}(s)$, which are close to the usual
strong couplant $a_s(Q^2)$ in the limit of large  $Q^2$ values and completely different from $a_s(Q^2)$ for small  $Q^2$ values,
i.e. for $Q^2 \sim \Lambda^2$.

A further APT development  is the so-called fractional APT (FAPT) \cite{BMS1,Bakulev:2006ex,Bakulev:2010gm}, which extends the construction principles
described above to PT series, starting from non-integer powers of the couplant. In the framework of QFT, such series arise for quantities that have
non-zero anomalous dimensions.
Compact expressions for quantities within the FAPT framework were obtained mainly in LO, but this approach was also used in higher orders,
mainly by re-expanding the corresponding couplants in powers of the LO couplant, as well as using some approximations.

In this short paper, we give an overview of the main properties of
MA couplants in the FAPT framework, obtained in Refs.  \cite{Kotikov:2022sos,KoZe23}
using the so-called $1/L$-expansion. Note that for an ordinary couplant, this expansion is applicable only for large $Q^2$ values, i.e. for $Q^2>>\Lambda^2$.
However, as shown in  \cite{Kotikov:2022sos,KoZe23},  the situation is quite different in the case of analytic couplants,
and this $1/L$-expansion is applicable for all values of the argument. This is due to the fact that the non-leading expansion corrections vanish not only
at $Q^2 \to \infty$, but also at $Q^2 \to 0$,
\footnote{The absence of high-order corrections for $Q^2 \to 0$ was also discussed in Refs. \cite{ShS,MSS}.}
which leads only to nonzero (small) corrections in the region $Q^2 \sim \Lambda^2$. 

Below we consider the representations for the MA couplants and their (fractional) derivatives obtained in  \cite{Kotikov:2022sos,KoZe23} and valid in
principle in any PT order. However, in order to avoid cumbersome formulas, but at the same time to show the main features of the approach
obtained in  \cite{Kotikov:2022sos,KoZe23}, we confine ourselves to considering only the first three PT orders.

\section{Strong couplant}
\label{strong}

As shown in the Introduction, the strong couplant $a_s(Q^2)$ obeys the renormalized group equation (\ref{RenGro}).
When $Q^2>>\Lambda^2$, Eq. (\ref{RenGro})  can be solved by iterations in the form of a $1/L$-expansion
(we give the first three terms of the expansion in accordance with the reasoning in the introduction),
which can be represented in the following compact form
\be
a^{(1)}_{s,0}(Q^2) = \frac{1}{L_0},~~
a^{(i+1)}_{s,i}(Q^2) = 
a^{(1)}_{s,i}(Q^2) + \sum_{m=2}^i \, \delta^{(m)}_{s,i}(Q^2)
\,,~~(i=0,1,2,...)\,,
\label{as}
\ee
where
\be
L_k=\ln t_k,~~t_k=\frac{1}{z_k}=\frac{Q^2}{\Lambda_k^2}\,.
\label{L}
\ee

The corrections $\delta^{(m)}_{s,k}(Q^2)$ are represented as follows
\be
\delta^{(2)}_{s,k}(Q^2) = - \frac{b_1\ln L_k}{L_k^2} ,~~
\delta^{(3)}_{s,k}(Q^2) =  \frac{1}{L_k^3} \, \Bigl[b_1^2(\ln^2 L_k-\ln L_k-1)+b_2\Bigr]\,.
\label{ds}
\ee

As shown in Eqs. (\ref{as}) and (\ref{ds}), in any PT
order, the couplant $\ar(Q^2)$ contains its dimensional transmutation parameter
$\Lambda$, which is related to the normalization of $\alpha_s(M_Z^2)$,
where $\alpha_s(M_Z)=0.1176$ in  PDG20 \cite{PDG20}.\\

{\bf $f$-dependence of the couplant $\ar(Q^2)$.}~~
The coefficients $\beta_i$ (\ref{beta_i}) depend on the number $f$ of active quarks
that change the couplant $\ar(Q^2)$ at  thresholds $Q^2_f \sim m^2_f$, where some  the additional quark comes enters the game $Q^2 > Q^2_f$.
Here $m_f$ is the $\overline{MS}$  mass of the $f$ quark, e.g.,
$m_b=4.18 +0.003-0.002$ GeV and $m_c=1.27 \pm 0.02$ GeV from PDG20 \cite{PDG20}.
\footnote{Strictly speaking, the quark masses in the $\overline{MS}$ scheme depend on $Q^2$ and $m_f=m_f(Q^2=m_f^2)$. The $Q^2$-dependence is rather slow and
  will not be discussed in this paper.}
Thus, the couplant $a_s$ depends on $f$, and this $f$-dependence can be taken into account in $\Lambda$, i.e. it is $\Lambda^f$ that contributes to the above
Eqs. (\ref{RenGro}) and (\ref{as}).

Relationships between $\Lambda_{i}^{f}$ and $\Lambda_{i}^{f-1}$, i.e.
the so-called matching conditions
between $a_s(f,Q_f^2)$ and $a_s(f-1,Q_f^2)$
are known up to the four-loop order \cite{Chetyrkin:2005ia} in the $\overline{MS}$ scheme and  usually are used
for $Q_f^2=m_f^2$, where these relations have the simplest form (see e.g. \cite{Enterria} for a recent review).

Here we will not consider the $f$-dependence of $\Lambda_{i}^{f}$ and $a_s(f,M_Z^2)$, since we mainly consider the range of small $Q^2$ values and therefore use
$\Lambda_{i}^{f=3}$.

\begin{figure}[!htb]
\centering
\includegraphics[width=0.58\textwidth]{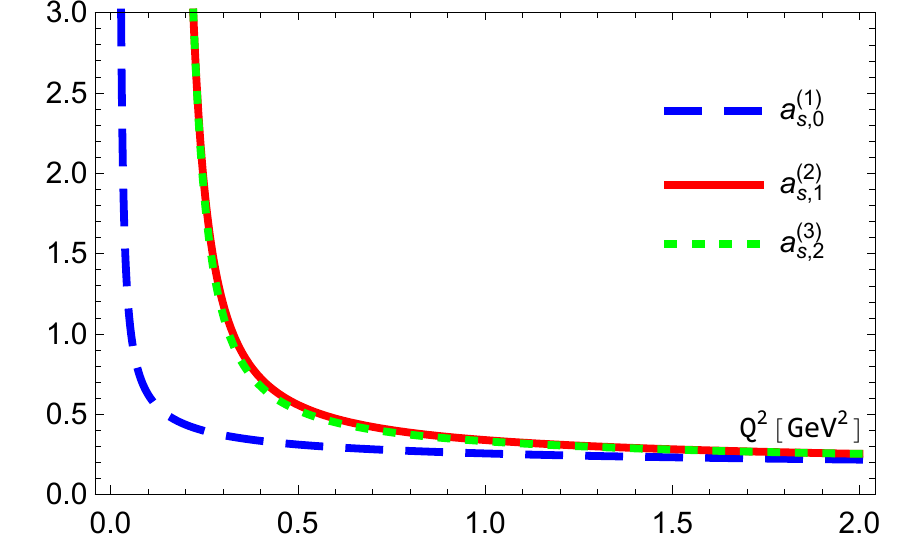}
\caption{\label{fig:as1352}
  The results for $a^{(i+1)}_{s,i}(Q^2)$ with $i=0,1,2$.
}
\end{figure}

On Fig. \ref{fig:as1352} one can see that the strong couplants $a^{(i+1)}_{s,i}(Q^2)$ become singular at $Q^2=\Lambda_i^2$.
The values of $\Lambda_0$ and $\Lambda_i$ $(i\geq 1)$ are very different.
We use results taken from a recent Ref. \cite{Chen:2021tjz}, where $\Lambda_i^{f=3}$
$(i=0,1,2)$ were obtained in the following form
\be
\Lambda_0^{f=3}=142~~ \mbox{MeV},~~\Lambda_1^{f=3}=367~~ \mbox{MeV},~~\Lambda_2^{f=3}=324~~ \mbox{MeV}
\,.
\label{Lambdas}
\ee

\section{Fractional derivatives}

Following \cite{Cvetic:2006mk,Cvetic:2006gc},
we introduce the derivatives (in the $(i)$-order of of PT)
\be
\tilde{a}^{(i)}_{n+1}(Q^2)=\frac{(-1)^n}{n!} \, \frac{d^n a^{(i)}_s(Q^2)}{(dL)^n} \, ,
\label{tan+1}
\ee
which are very convenient in the case of the analytical QCD (see, e.g., \cite{Kotikov:2022JETP}).

The series of derivatives $\tilde{a}_{n}(Q^2)$ can successfully replace the corresponding series of $\ar$-degrees. Indeed, each
the derivative reduces the $\ar$ degree, but is accompanied by an additional $\beta$-function $\sim \ar^2$.
Thus, each application of a derivative yields an additional $\ar$, and thus  indeed possible to use series of derivatives instead of
series of $\ar$-powers.

In LO, the series of derivatives $\tilde{a}_{n}(Q^2)$ are exactly the same as $\ar^{n}$. Beyond LO, the relationship between $\tilde{a}_{n}(Q^2)$
and $\ar^{n}$ was established in \cite{Cvetic:2006gc,Cvetic:2010di} and extended to fractional cases, where $n \to$ is a non-integer $\nu $,
in Ref. \cite{GCAK}.

Now consider the $1/L$-expansion of $\tilde{a}^{(k)}_{\nu}(Q^2)$. We can raise the $\nu$-power of the results (\ref{as}) and (\ref{ds}) and then
restore $\tilde{a}^{(k)}_{ \nu}(Q^ 2)$ using the relations between $\tilde{a}_{\nu}$ and $\ar^{\nu}$ obtained in \cite{GCAK}.
This operation is carried out in detail in Appendix B to \cite{Kotikov:2022sos} (see also  Appendix A to \cite{Kotikov:2022vnx}).
Here we present only the final results, which have the following form
\footnote{The expansion (\ref{tdmp1N}) is similar to those used in Refs. \cite{BMS1,Bakulev:2006ex} for the expansion of
  ${\bigl({a}^{(i+1)} _{s,i}(Q^2)\bigr)}^ {\nu}$ in terms of powers of $a^{(1)}_{s,i}(Q^2)$.}:
\bea
\z\tilde{a}^{(1)}_{\nu,0}(Q^2)={\bigl(a^{(1)}_{s,0}(Q^2)\bigr)}^{\nu} = \frac{1}{L_0^{\nu}},~
\tilde{a}^{(i+1)}_{\nu,i}(Q^2)=\tilde{a}^{(1)}_{\nu,i}(Q^2) + \sum_{m=1}^{i}\, C_m^{\nu+m}\, \tilde{\delta}^{(m+1)}_{\nu,i}(Q^2),~~\nonumber\\
\z\tilde{\delta}^{(m+1)}_{\nu,i}(Q^2)=
\hat{R}_m \, \frac{1}{L_i^{\nu+m}},~~C_m^{\nu+m}=\frac{\Gamma(\nu+m)}{m!\Gamma(\nu)}\,,
\label{tdmp1N}
\eea
where
\be
\hat{R}_1=b_1 \Bigl[\hat{Z}_1(\nu)+ \frac{d}{d\nu}\Bigr],~~
\hat{R}_2=b_2 + b_1^2 \Bigl[\frac{d^2}{(d\nu)^2} +2 \hat{Z}_1(\nu+1)\frac{d}{d\nu} + \hat{Z}_2(\nu+1 )\Bigr]
\, 
\label{hR_i}
\ee
and $\hat{Z}_j(\nu)$ $(j=1,2)$ are combinations of the Euler $\Psi$-functions and their derivatives.

The representation (\ref{tdmp1N}) of the $\tilde{\delta}^{(m+1)}_{\nu,i}(Q^2)$ corrections as $\hat{R} _m$-operators is very important and allows us
to  similarly present high-order results for the ($1/L$-expansion) of analytic couplants.

\section{MA coupling}

We  first show the LO results, and then go beyond LO following our results (\ref{tdmp1N}) for the ordinary  strong couplant obtained in the previous section.\\

{\bf LO.}~~
The LO MA couplant $A^{(1)}_{{\rm MA},\nu,0}$
has the following form  \cite{BMS1}
\be
A^{(1)}_{{\rm MA},\nu,0}(Q^2) = {\left( a^{(1)}_{\nu,0}(Q^2)\right)}^{\nu} - \frac{{\rm Li}_{1-\nu}(z_0)}{\Gamma(\nu)}=
\frac{1}{L_0^{\nu}}- \frac{{\rm Li}_{1-\nu}(z_0)}{\Gamma(\nu)} \equiv \frac{1}{L_0^{\nu}}-\Delta^{(1)}_{\nu,0}\,,
\label{tAMAnu}
\ee
where
\be
   {\rm Li}_{\nu}(z)=\sum_{m=1}^{\infty} \, \frac{z^m}{m^{\nu}}=  \frac{z}{\Gamma(\nu)} \int_0^{\infty} 
\frac{ dt \; t^{\nu -1} }{(e^t - z)}
   \label{Linu}
\ee
is the Polylogarithm.

The LO MA couplant $U^{(1)}_{{\rm MA},\nu,0}$ in the Minkowski space
has the
form \cite{Bakulev:2006ex}
\be
U^{(1)}_{{\rm MA},\nu,0}({\rm s})
=\frac{\sin[(\nu-1)\,g_0(s)]
}{\pi(\nu-1)(\pi^2+ L^2_{s,0})^{(\nu-1)/2}}
  \, ,\, (\nu>0)\, ,
\label{mainexpr}
\ee
where
\be
L_{s,i}=\ln\dfrac{s}{\Lambda_i^2},~~g_i(s)= \arccos\left(\frac{L_{s,i}}{\sqrt{\pi^2+ L^2_{s,i}}}\right) \, .
\label{Ls}
\ee

For $\nu=1$ we recover the famous Shirkov-Solovtsov results  \cite{ShS}:
\be
\hspace{-0.5cm} A^{(1)}_{\rm MA,0}(Q^2) \equiv A^{(1)}_{\rm MA,\nu=1,0}(Q^2)
=\frac{1}{L_0}- \frac{z_0}{1-z_0},~ U^{(1)}_{\rm MA,0}(Q^2) \equiv U^{(1)}_{\rm MA,\nu=1,0}({\rm s})=\frac{g_0(s)}{\pi}\, .
\label{tAM1}
\ee
Note that the result (\ref{tAM1}) can be taken directly for the integral forms (\ref{disp_MA_LO}) and (\ref{disp_MAt_LO}), as it was in Ref. \cite{ShS}.\\

{\bf Beyond LO.}
Following Eqs. (\ref{tAMAnu}) and (\ref{mainexpr}) for the LO analytic couplants, 
we consider 
the derivatives of the
MA couplants, as
\be
\tilde{A}_{{\rm MA},n+1}(Q^2)=\frac{(-1)^n}{
  n!} \, \frac{d^n A_{\rm MA}(Q^2)}{(dL)^n},~~\tilde{U}_{{\rm MA},n+1}(Q^2)=\frac{(-1)^n}{
  n!} \, \frac{d^n U_{\rm MA}(s)}{(dL_s)^n} \, .
\label{tanMA+1}
\ee

By analogy with ordinary couplant,
using the results (\ref{tdmp1N})
we have for MA analytic couplants $\tilde{A}^{(i+1)}_{{\rm MA},\nu,i}$ and $\tilde{U}^{(i+1)}_{{\rm MA},\nu,i}$ the following expressions:
\bea
&&\tilde{A}^{(i+1)}_{{\rm MA},\nu,i}(Q^2) = \tilde{A}^{(1)}_{{\rm MA},\nu,i}(Q^2) + \sum_{m=1}^{i}  \, C^{\nu+m}_m \tilde{\delta}^{(m+1)}_{{\rm A},\nu,i}(Q^2), \nonumber \\
&&\tilde{U}^{(i+1)}_{{\rm MA},\nu,i}(s) = \tilde{U}^{(1)}_{{\rm MA},\nu,i}(s) + \sum_{m=1}^{i}  \, C^{\nu+m}_m \tilde{\delta}^{(m+1)}_{{\rm U},\nu,i}(s),
\label{tAiman}
\eea
where  $\tilde{A}^{(1)}_{{\rm MA},\nu,i}$ and $\tilde{U}^{(1)}_{{\rm MA},\nu,i}$ are given in Eqs.  (\ref{tAMAnu}) and (\ref{mainexpr}), respectively, and
\be
\tilde{\delta}^{(m+1)}_{{\rm A},\nu,i}(Q^2)= \tilde{\delta}^{(m+1)}_{\nu,i}(Q^2) -  \hat{R}_m \left( \frac{{\rm Li}_{-\nu-m+1}(z_i)}{\Gamma(\nu+m)}\right),~
\tilde{\delta}^{(m+1)}_{{\rm U},\nu,i}(s)= \hat{R}_m \Bigl(\tilde{U}^{(1)}_{{\rm MA},\nu+m,i}(s)\Bigr)\,.
\label{tdAman}
\ee
and $\tilde{\delta}^{(m+1)}_{\nu,i}(Q^2)$ and $\hat{R}_m$
are given in Eqs. (\ref{tdmp1N}) and (\ref{hR_i}), respectively.\\

The analytical results for the MA analytic couplants $\tilde{A}^{(i+1)}_{{\rm MA},\nu,i}$ and $\tilde{U}^{(i+1)}_{{\rm MA},\nu,i}$
can be found in Refs. \cite{Kotikov:2022sos} and \cite{KoZe23}, respectively. Here we present only the results for the case $\nu=1$:
\bea
&&A^{(i+1)}_{{\rm MA},i}(Q^2)\equiv \tilde{A}^{(i+1)}_{{\rm MA},\nu=1,i}(Q^2) = A^{(1)}_{{\rm MA},i}(Q^2) + \sum_{m=1}^{i}  \, \tilde{\delta}^{(m+1)}_{{\rm A},\nu=1,i}(Q^2),~~\nonumber \\
&&U^{(i+1)}_{{\rm MA},i}(s)\equiv \tilde{U}^{(i+1)}_{{\rm MA},\nu=1,i}(s) = U^{(1)}_{{\rm MA},i}(s) + \sum_{m=1}^{i}  \, \tilde{\delta}^{(m+1)}_{{\rm U},\nu=1,i}(s)
\label{tAiman.1}
\eea
where $A^{(1)}_{{\rm MA},i}(Q^2)$ and $ U^{(1)}_{{\rm MA},i}(s)$ are shown in Eq. (\ref{tAM1}) and
\bea
\z  \tilde{\delta}^{(m+1)}_{{\rm A},\nu=1,i}(Q^2)
= \tilde{\delta}^{(m+1)}_{\nu=1,i}(Q^2)- \frac{P_{m,1}(z_i)}{m!} \, ,\nonumber\\
\z  \tilde{\delta}^{(2)}_{{\rm A},\nu=1,i}({\rm s})= \frac{b_1}{\pi(\pi^2+L_{s,i}^2)^{1/2}}\Bigl\{g_i\cos(g_i)-\Bigl[1+G_i\Bigr]\sin(g_i)\Bigr\},~~
\nonumber\\
\z  \tilde{\delta}^{(3)}_{{\rm U},\nu=1,i}({\rm s})=\frac{1}{2\pi(\pi^2+L_s^2)}\Biggl(b_2\sin(2g_i) +b_1^2\Bigl[G_i^2-g_i^2-1\Bigr]\sin(2g_i)\Biggr)
\label{tdAmanA}
\eea
with 
\bea
\z G_i({\rm s})=\frac{1}{2}\,\ln\left(\pi^2+L_{s,i}^2\right),~~P_{1,\nu}(z)=b_1\Bigl[\overline{\gamma}_{\rm E}
       {\rm Li}_{-\nu}(z)+{\rm Li}_{-\nu,1}(z)\Bigr],~~\overline{\gamma}_{\rm E}=\gamma_{\rm E}-1,~~ \nonumber \\
\z P_{2,\nu}(z)=b_2 \,{\rm Li}_{-\nu-1}(z) + b_1^2\Bigl[{\rm Li}_{-\nu-1,2}(z) + 2\overline{\gamma}_{\rm E}
  {\rm Li}_{-\nu-1,1}(z)
  +  \Bigl(\overline{\gamma}^2_{\rm E}-
  \zeta_2\Bigr) \, {\rm Li}_{-\nu-1}(z) \Bigr]\,,
\label{Pkz}
\eea
Euler constant $\gamma_{\rm E}$ and
\be
 {\rm Li}_{n,m}(z)= \sum_{m=1} \, \frac{\ln^k m}{m^n},~~
    {\rm Li}_{-1}(z)= \frac{z}{(1-z)^2},~~{\rm Li}_{-2}(z)= \frac{z(1+z)}{(1-z)^3}
    \, .
\label{Lii.1}
\ee

\begin{figure}[!htb]
\centering
\includegraphics[width=0.58\textwidth]{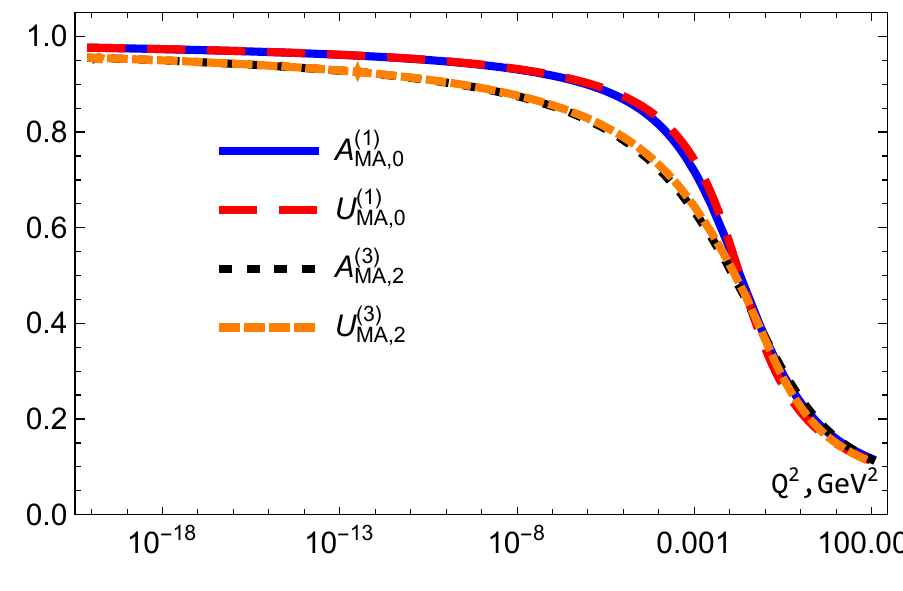}
    \caption{\label{fig:A123}
      The results for $A^{(i+1)}_{\rm MA,i}(Q^2)$ and $U^{(i+1)}_{\rm MA,i}(Q^2)$ with $i=0,2$.}
\end{figure}

On Fig. \ref{fig:A123} we see that $A^{(i+1)}_{\rm MA,i}(Q^2)$ and $U^{(i +1) }_{\rm MA,i}(Q^2)$ are very close to each other for $i=0$ and $i=2$.
The differences between the L0 and NNLO results are nonzero only for $Q^2 \sim \Lambda^2$.

\section{Conclusions}

In this short paper, we have demonstrated the results obtained in our recent papers \cite{Kotikov:2022sos,KoZe23}. In particular,
Ref.  \cite{Kotikov:2022sos} contains $1/L$-expansions of $\nu$-derivatives of the strong couplant $a_s$ expressed as combinations of
the $\hat{R}_m$ (\ref{hR_i}) operators applied to the LO couplant $a_s^{(1)}$.
Using the same operators to $\nu$-derivatives of LO MA couplants $A_{\rm MA}^{(1)}$ and $U_{\rm MA}^{(1)}$, various
representations were obtained for $\nu$-derivatives of MA couplants,
i.e. $\tilde{A}_{\rm MA,\nu}^{(i)}$ and  $\tilde{U}_{\rm MA,\nu}^{(i)}$ in each $i$-order of PT.
All results are presented in \cite{Kotikov:2022sos,KoZe23} up to the 5th order of PT,
where the corresponding QCD $\beta$-function coefficients are well known
(see \cite{Baikov:2016tgj}). In this paper, we have limited ourselves to the first three orders in order to exclude the most cumbersome
results obtained for the last two PT orders.

High-order corrections are negligible in both asymptotics: $Q^2 \to 0$ and $Q^2 \to \infty$, and are nonzero in a neighborhood of the point $Q^2 =\Lambda^2$.
Thus, in fact, they  represent only minor corrections to LO MA couplants $A_{\rm MA}^{(1)}(Q^2)$ and $U_{\rm MA}^{(1)}(Q^2)$.
This proves the possibility of expansions of high-order couplants $A_{\rm MA}^{(i)}(Q^2)$ and $U_{\rm MA}^{(i)}(Q^2)$ via the
LO couplants $A_{\rm MA}^{(1)}(Q^2)$ and $U_{\rm MA}^{(1)}(Q^2)$, which was done in
Ref. \cite{Bakulev:2010gm}.\\


{\bf Acknowledgments}~
This work was supported in part by the Foundation for the Advancement
of Theoretical Physics and Mathematics “BASIS”.
One of us (A.V.K.) thanks the Organizing Committee of the XXIV International Seminar on High Energy Physics "From quarks to
galaxies: clearing up the dark sides" (November 22-24, Protvino, Russia)
for invitation.

\end{document}